\documentclass [prb,superscriptaddress, showpacs, twocolumn]{revtex4-1}
\usepackage{graphicx}
\usepackage{dcolumn}
\usepackage{bm}

\makeatletter

\newcommand{\Rmnum}[1]{\expandafter\@slowromancap\romannumeral #1@}
\makeatother

\begin{document}

\title{Polar and antipolar polymorphs of metastable perovskite BiFe$_{0.5}$Sc$_{0.5}$O$_3$}

\author{D. D. Khalyavin}
\email{email: dmitry.khalyavin@stfc.ac.uk}
\affiliation{ISIS facility, STFC, Rutherford Appleton Laboratory, Chilton, Didcot, Oxfordshire, OX11-0QX,UK}
\author{A. N. Salak}
\email{email: salak@ua.pt}
\affiliation{Department of Materials and Ceramic Engineering/CICECO, University of Aveiro, 3810-193 Aveiro, Portugal}
\author{N. M. Olekhnovich}
\affiliation{Scientific-Practical Materials Research Centre of NAS of Belarus, Minsk, 220072, Belarus}
\author{A. V. Pushkarev}
\affiliation{Scientific-Practical Materials Research Centre of NAS of Belarus, Minsk, 220072, Belarus}
\author{Yu. V. Radyush}
\affiliation{Scientific-Practical Materials Research Centre of NAS of Belarus, Minsk, 220072, Belarus}
\author{P. Manuel}
\affiliation{ISIS facility, STFC, Rutherford Appleton Laboratory, Chilton, Didcot, Oxfordshire, OX11-0QX,UK}
\author{I. P. Raevski}
\affiliation{Faculty of Physics and Research Institute of Physics, Southern Federal University, 344090, Rostov-on-Don, Russia}
\author{M. L. Zheludkevich}
\affiliation{Department of Materials and Ceramic Engineering/CICECO, University of Aveiro, 3810-193 Aveiro, Portugal}
\affiliation{MagIC, Institute of Materials Research, Helmholtz-Zentrum Geesthacht, 21502 Geesthacht, Germany}
\author{M. G. S. Ferreira}
\affiliation{Department of Materials and Ceramic Engineering/CICECO, University of Aveiro, 3810-193 Aveiro, Portugal}
\date{\today}

\begin{abstract}
A metastable perovskite BiFe$_{0.5}$Sc$_{0.5}$O$_3$ synthesized under high-pressure (6 GPa) and high-temperature (1500 K) conditions was obtained in two different polymorphs, antipolar $Pnma$ and polar $Ima2$, through an irreversible behaviour under a heating/cooling thermal cycling. The $Ima2$ phase represents an original type of a canted ferroelectric structure where Bi$^{3+}$ cations exhibit both polar and antipolar displacements along the orthogonal $[110]_p$ and $[1\bar{1}0]_p$ pseudocubic directions, respectively, and are combined with antiphase octahedral tilting about the polar axis. Both the $Pnma$ and $Ima2$ structural modifications exhibit a long-range antiferromagnetic ordering with a weak-ferromagnetic component below $T_N\sim220$ K. Analysis of the coupling between the dipole, magnetic and elastic order parameters based on a general phenomenological approach revealed that the weak-ferromagnetism in both phases is mainly caused by the presence of the antiphase octahedral tilting whose axial nature directly represents the relevant part of Dzyaloshinskii vector. The magnetoelectric contribution to the spontaneous magnetization allowed in the polar $Ima2$ phase is described by a fifth-degree free-energy invariant and is expected to be small.
\end{abstract}

\pacs{75.25.-j}

\maketitle

\section{Introduction}
\indent Materials which combine two and more ferroic order parameters (multiferroics) have attracted extensive research interest in recent several years due to possibility of effective cross-control of their properties, providing avenue for designing new multifunctional electronic devices.\cite{ref:1,ref:2,ref:3} In this respect, a coexistence of macroscopic polarization and magnetization are of particular importance, since their interference may offer an opportunity to control dielectric properties by magnetic field and vice versa. The coexistence of ferroelectricity and spontaneous magnetization is however an extremely rare phenomenon due to the mutually exclusive nature of these order parameters.\cite{ref:4} If one adds on top of that the requirement of room temperature functionality, we are left with the single example of bismuth ferrite (BiFeO$_3$). The transitions to the ferroelectric $R3c$ and magnetic phases in this compound take place at $\sim 1100$ K and $\sim 640$ K, respectively.\cite{ref:5,ref:6} In a first approximation, the magnetic order is antiferromagnetic but a spin-canting is allowed by symmetry, if the spins are confined to be within the $(ab)$ plane. The origin of the weak-ferromagnetism in BiFeO$_3$ has been assigned by Kadomtseva et al.\cite{ref:7} to "magnetoelectric mechanism", where polarization acts as internal electric fieled generating the magnetization through magnetoelectric coupling. On the other hand, Ederer and Spaldin\cite{ref:8} by means of density functional calculations found that the tilting of oxygen octahedra, which also presents in the polar structure of BiFeO$_3$, is the relevant distortion to induce the spin-canting. The phenomenological approach adopted by Kadomtseva et al.\cite{ref:7} used the "minimal" centrosymmetric supergroup $R\bar{3}c1'$ as the parent symmetry to evaluate the form of the free-energy decomposition. This approach is strongly restricted; in order to exploit fully the advantage of this symmetry-based method, the cubic $Pm\bar{3}m1'$ space group should be used instead. The ferromagnetic and antiferromagnetic order parameters are decoupled (transform accordingly different space group representations) in respect of this cubic symmetry and therefore the necessary conditions to combine them into the appropriate free-energy invariant can be obtained in the most general way.\\
\indent In fact, the polar nature of the crystal structure of BiFeO$_3$, in the magnetically ordered phase, induces a slow spin-rotation due to the relativistic part of the exchange interactions, forming a long-period cycloid and averaging the local weak-ferromagnetic components to zero.\cite{ref:9} The phase with a non-zero macroscopic magnetization however can be stabilized by external magnetic field destroying the spin-cycloid.\cite{ref:7,ref:10} Taking into account this drawback, a number of attempts has been performed to stabilize the zero-field polar weak-ferromagnetic state in BiFeO$_3$ through a chemical doping and thin-film strain engineering.\\
\indent From the crystal chemistry point of view, BiFeO$_3$ belongs to the family of perovskite, $AB$O$_3$, materials whose ability to accommodate isovalent and heterovalent substitutes in both $A$- and $B$- structural positions is well known and widely used for properties control and improvement.\cite{ref:11} One of the main doping-strategies applied to BiFeO$_3$ is focused on lanthanum\cite{ref:12,ref:13,ref:14} or rare-earth substitution of Bi.\cite{ref:15,ref:16,ref:17,ref:18,ref:19} This allows to keep the charge balance without changing the charge-state of the Fe-sublattice or oxygen stoichiometry. In such manner, rich "temperature-composition" phase diagrams have been revealed with a variety of compositionally driven structural phase transitions.\cite{ref:12} Some of the phases demonstrate complex superstructures related to antiferroelectric and even incommensurate displacements of Bi. The presence of the ferroelectrically inactive cations in the Bi-sublattice quickly destroys the long-range dipole ordering making the compositions irrelevant to explore the relations between dielectric and magnetic properties. Studies of the systems derived from BiFeO$_3$ by means of substitutions in the Fe-sublattice are rather scarce; mainly because of a limited solubility even in the case of cations which are close to octahedrally coordinated Fe$^{3+}$ in terms of ionic size. High-pressure (HP) synthesis technique makes it possible to extend the solubility and thereby the perovskite phase existence ranges. In many cases, metastable perovskite phases can be stabilized by quenching under pressure. These phases then can be studied using the same facilities applicable to the stable phases over wide temperature range at ambient pressure. However, this technique is not easily available and few HP perovskite systems based on BiFeO$_3$ have been reported so far. The entire series of solid solutions have been obtained for the BiFeO$_3$-BiMnO$_3$ system only;\cite{ref:20} in other cases, compositions with the particular ratio Fe$^{3+}$/$B^{3+}$=1:1, where $B^{3+}$ = Cr and Mn were merely studied.\cite{ref:21,ref:22}\\
\indent It should be noticed that preparation and investigation of metastable perovskites (both particular compounds and solid solutions) are of a great interest in several respects. First, such perovskites can be stabilized in new crystal structures and hence offer novel properties and effects. Second, analysis of crystal structure and physical characteristics of a metastable phase in comparison with those of the phases which are stable at ambient pressure is a way to reveal and/or clarify the regularities and phenomena which have not been noticed before.\cite{ref:23} Third, some perovskite phases (including Bi-containing ones), which can be obtained in a bulk form through a HP synthesis only,\cite{ref:24} turned out to be grown as epitaxial films using rather conventional methods.\cite{ref:25}\\
\indent In this work, a "BiFe$_{0.5}$Sc$_{0.5}$O$_3$" perovskite has been the focus of our research. This composition is actually the equimolar solid solution in the BiFeO$_3$ - BiScO$_3$ system. Bismuth scandate perovskite is a HP-phase characterized by the monoclinic $C2/c$ space group.\cite{ref:26} Considering the different symmetries of the parent compounds and the considerable ionic size difference between Fe$^{3+}$ and Sc$^{3+}$, one can expect a variety of perovskite phases in this system. Here, we found that BiFe$_{0.5}$Sc$_{0.5}$O$_3$ quenched under pressure is antiferroelectric which can be turned into a polar phase by irreversible heating/cooling thermal cycling. This polar modification is a new type of distorted perovskite structure which combines spontaneous polarization and magnetization at high temperature $\sim$220 K. The system has a great potential to bring the multiferroic properties up to room temperature by slight modification of the Fe/Sc ratio and is therefore very attractive to be explored in a strain-stabilized epitaxial form. Moreover, taking into account the variety of quasi-degenerate phases recently discovered theoretically in the energy landscape of BiFeO$_3$,\cite{ref:27} the temperature-induced irreversible behaviour observed in the present work can hopefully be found in other metastable perovskites derived from BiFeO$_3$. 

\section{Experimental section}
\indent High-purity oxides Bi$_2$O$_3$, Fe$_2$O$_3$, and Sc$_2$O$_3$ were used as starting reagents to prepare the composition BiFe$_{0.5}$Sc$_{0.5}$O$_3$. Previously calcined oxides were mixed in the stoichiometric ratio, ball-milled in acetone, dried, and pressed into pellets. The pellets were heated in a closed alumina crucible at 1140 K for 10 min and then quenched down to room temperature. The obtained material served as a precursor for the HP synthesis. High pressure was generated using an anvil press DO-138A with a press capacity up to 6300 kN (SPMRC, Minsk). In order to avoid penetration of graphite from the tubular heater to the sample a protective screen of molybdenum foil was used. The samples were synthesized at 6 GPa and 1500 K for 1-3 min.\\
\indent An X-ray diffraction (XRD) study of the powders was performed using a PANalytical X'Pert MPD PRO diffrac-tometer (Ni-filtered Cu K$\alpha $ radiation, tube power 45 kV, 40 mA; PIXEL detector, and the exposition corresponded to about 2 s per step of 0.02$^{\circ }$ over the angular range 15 - 100$^{\circ }$) at room temperature. In situ XRD measurements were conducted in an Anton Paar High-Temperature Chamber (HTK 16N) in a temperature range between 300 and 820 K.\\
\indent Neutron powder diffraction data were collected at the ISIS pulsed neutron and muon facility of the Rutherford Appleton Laboratory (UK), on the WISH diffractometer located at the second target station.\cite{ref:28} The samples ($\sim$25 mg each) were loaded into cylindrical 3 mm diameter vanadium cans and measured in the temperature range of 1.5 - 300 K (step 30 K, exposition time 2h) using an Oxford Instrument Cryostat. Rietveld refinements of the crystal and magnetic structures were performed using FullProf program\cite{ref:29} against the data measured in detector banks at average 2$\theta $ values of 58$^{\circ }$, 90$^{\circ }$, 122$^{\circ }$, and 154$^{\circ }$, each covering 32$^{\circ }$ of the scattering plane.\\
\indent Magnetization data were measured using a supercon-ducting quantum interference device (SQUID) magne-tometer (Quantum Design MPMS).\\
\indent The sample was not annealed before the in situ XRD study. In the first cycle, the sample was heated to 370 K and then temperature was increased up to 820 K with a step of 50 K. At each temperature point, the sample was held for 1 h before the XRD measurement. After 820 K, the sample was stepwise cooled down to room tempera-ture with measurements at 770 to 370 K every 100 K without any holding. In the second cycle, the XRD measurements on the same sample were conducted without holding on both heating and cooling.\\
\indent The product of the 10-min reaction at ambient pressure (as described above) was found from the XRD study to be a mixture of phases, one of which was a rhombohedral perovskite phase. A more detailed analysis revealed that the perovskite phase was a BiFe$_{1-x}$Sc$_x$O$_3$ solid solution. Provided that a dependence of the reduced unit cell parameter on $x$ is linear, amount of BiScO$_3$ in the solution was estimated to be $\sim$15 at.$\%$. Besides, Bi$_{19}$ScO$_{30}$, which is a composition based on either $\beta$-Bi$_2$O$3$ or $\gamma $-Bi$_2$O$_3$ containing about 10 at.$\%$ of Sc$_2$O$_3$,\cite{ref:30} and some amount of unreacted scandium oxide were revealed. Double homogenization, increase of the reaction time, and variation of the heating/cooling rates resulted in changes in neither quantitative ratio nor qualitative content of the observed phases.\\
\indent In order to estimate the stability limit of the HP-synthesized BiFe$_{0.5}$Sc$_{0.5}$O$_3$ perovskite, the samples were annealed in air at elevated temperatures followed by XRD study at room temperature. It was found that the perovskite phase is stable up to about 970 K. Annealing at higher temperatures resulted in appearance of the reflections associated with the Bi$_{19}$ScO$_{30}$ phase.

\section{Results}
\subsection{Crystal and magnetic structures of antipolar phase}
\begin{figure}[t]
\includegraphics[scale=1.0]{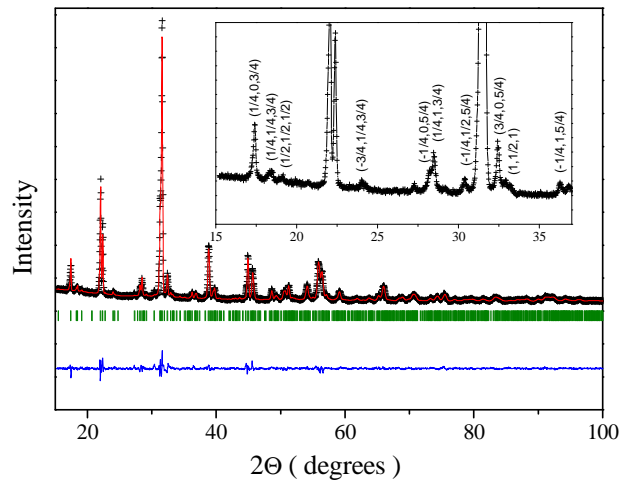}
\caption{(Color online) Rietveld refinement of X-ray diffraction data collected at 300 K on the as prepared BiFe$_{0.5}$Sc$_{0.5}$O$_3$ sample. The cross symbols and solid line (red) represent the experimental and calculated intensities, respectively, and the line below (blue) is the difference between them. Tick marks (green) indicate the positions of Bragg peaks in the $Pnma$ space group. Inset shows a zoomed low-angle region of the pattern with the pseudocubic indexation of the superstructure reflections.}
\label{Fig:F1}
\end{figure}

\begin{figure}[t]
\includegraphics[scale=1.0]{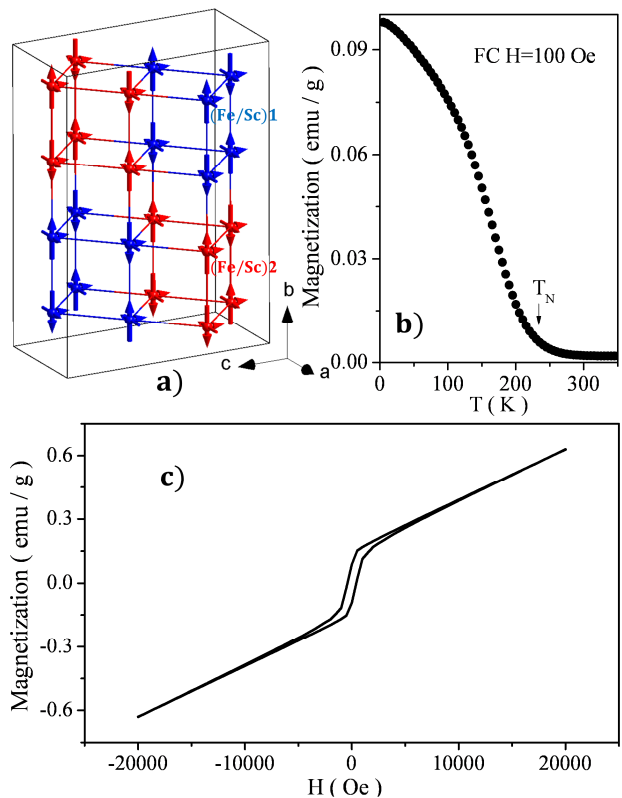}
\caption{(Color online) Magnetic structure of the as-prepared BiFe$_{0.5}$Sc$_{0.5}$O$_3$ sample with the $Pn'm'a$ magnetic space group. Two non-equivalent $8d$ positions of Fe/Sc are shown by different colour. The largest spin component $\sim$2.10(6) $\mu_B$ found in the neutron diffraction experiment is along the $b$-axis ($G_y$). The two orthogonal spin components, along the $a$- and $c$-axis allowed by the $Pn'm'a$ symmetry, are shown as well. (b) Magnetization as a function of temperature, measured under the magnetic field H=100 Oe after cooling under the same field. (c) Magnetization loop measured at 5 K after cooling in H=20000 Oe.}
\label{Fig:F4}
\end{figure}

The room temperature crystal structure of the as-prepared sample was determined from joint refinement of X-ray and time-of-flight neutron diffraction data (Fig. \ref{Fig:F1}). Several models were tested in the refinement procedure, deduced based on a comprehensive symmetry analysis.\cite{ref:31,ref:32} The initial analysis of the diffraction patterns revealed a clear splitting of the fundamental reflections and a complex superstructure both consistent with the orthorhombic-type $\sqrt{2}a_p\times 4a_p\times 2\sqrt{2}a_p$ supercell. The presence of reflections associated with ${\bm k}=1/2,1/2,1/2$ ($R$-point of symmetry), ${\bm k}=1/4,1/4,0$ ($\Sigma $-line of symmetry), ${\bm k}=1/4,1/4,1/4$ ($\Lambda$-line of symmetry), ${\bm k}=1/4,1/2,1/4$ ($S$-line of symmetry) and ${\bm k}=1/2,1/2,1/4$ ($T$-line of symmetry) modulation vectors, indicated extremely complex structural distortions. Taking into account the large number of arms in the each wave-vector star associated with the lines of symmetry, it was practically impossible to analyse explicitly all isotropy subgroups related to the corresponding reducible order parameters. Even selection of the minimal set of the propagation vectors which can form translational invariants with the rest modulations would not reduce the dimensionality of the coupled order parameters down to that solvable in a reasonable time. We, therefore, adopted a less general approach and made the analysis in several steps based on expected primary distortions typical of Bi-containing perovskites. These distortions are octahedral tilting (which is expected due to relatively small ionic size of the $A$-site cation) and ferroelectric/antiferroelectric displacements of Bi (due to stereochemically active electronic degree of freedom related to its lone pair). Initially, we combined only $R$- and $\Sigma $-type of distortions, namely $R_4^+\oplus\Sigma_2$ coupled order parameter, where $R_4^+$ and $\Sigma_2$ are the irreducible representations of the $Pm\bar{3}m$ space group transforming antiphase octahedral tilting and antiferroelectric displacements of Bi, respectively.$\cite{ref:33,ref:24}$ 
The superstructure reflections associated with these propagation vectors were found to be most intense, indicating their primary nature. For different directions of the order parameter in the reducible $R_4^+\oplus\Sigma_2$ representation space, sets of symmetry-adopted displacive modes were generated and checked directly in the refinement procedure. Then additional distortions were added to match the unit cell size of the resultant isotropy subgroup with the experimentally found $\sqrt{2}a_p\times 4a_p\times 2\sqrt{2}a_p$ supercell. The procedure resulted into two final candidates which provided near equally good refinement quality for the available diffraction data (Fig. \ref{Fig:F1}). Both candidates share identical $Pnma$ space group and are different in respect of a choice of the coordinate origin. It should be pointed out that these structures are non-equivalent in the sense that they do not represent different rotational/translational domains of the same structure and therefore are not related by any symmetry operation of the parent $Pm\bar{3}m$ space group. Moreover, in a general case, these structures have different free-energies. The final refinements were performed using the conventional "coordinate" approach and then the mode decomposition in respect of the symmetrized atomic displacements of the parent cubic perovskite structure has been done.\cite{ref:31,ref:32} 
\begin{figure}[t]
\includegraphics[scale=1.4]{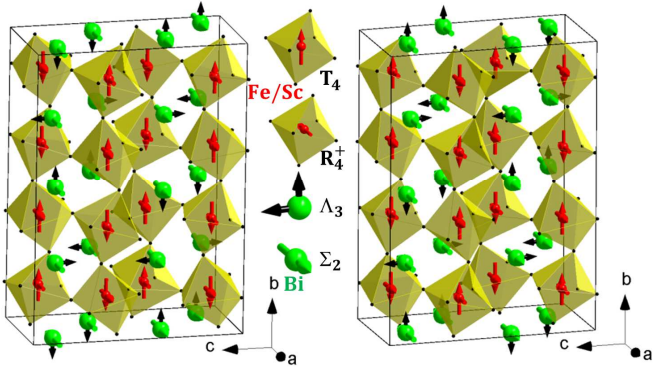}
\caption{(Color online) Schematic representation of the two possible antipolar subgroups for the as-prepared BiFe$_{0.5}$Sc$_{0.5}$O$_3$ sample with the $Pnma$ symmetry (lattice vectors with respect to the parent $Pm\bar{3}m$ group: ${\bm a_o}={\bm a_p}+{\bm c_p}, {\bm b_o}=4{\bm b_p}, {\bm c_o}=-2{\bm a_p}+2{\bm c_p}$) and the origin choice at ${\bm a_p}/2+{\bm b_p}/2$ (left) and ${\bm a_p}/2+3{\bm b_p}/2$ (right). Arrows, in the Bi and Fe/Sc positions, represent polar (displacements) and axial (octahedral rotations) vectors, respectively.}
\label{Fig:F2}
\end{figure}
\begin{figure}[t]
\includegraphics[scale=1.0]{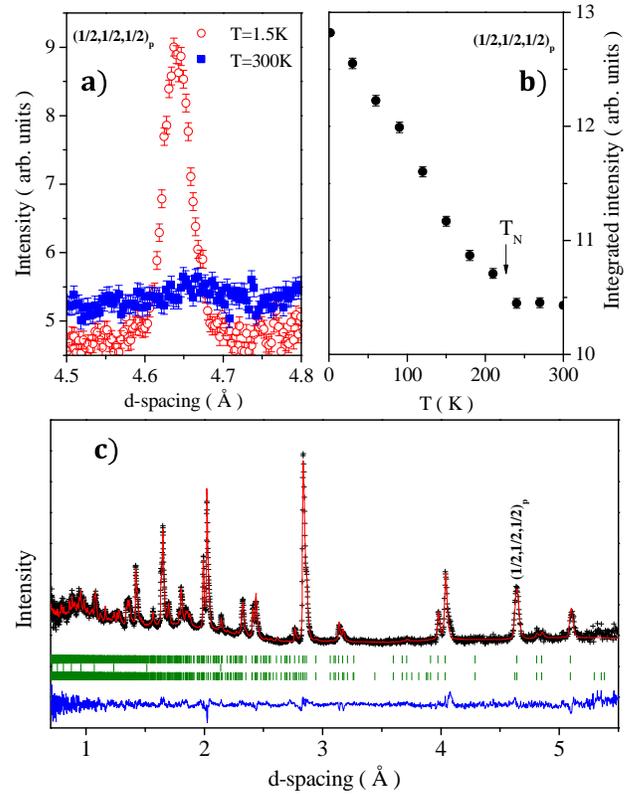}
\caption{(Color online) (a) Neutron diffraction patterns for the as-prepared BiFe$_{0.5}$Sc$_{0.5}$O$_3$ sample at the vicinity of the strongest $(1/2,1/2,1/2)_p$ magnetic peak collected above and below T$_N$. (b) Integrated intensity of this peak as a function of temperature. (c) Rietveld refinement of the neutron diffraction data collected at 1.5 K. The cross symbols and solid line (red) represent the experimental and calculated intensities, respectively, and the line below (blue) is the difference between them. Tick marks (green) indicate the positions of Bragg peaks: nuclear (top), vanadium can (middle) and magnetic ${\bm k}=0$ (bottom).}
\label{Fig:F3}
\end{figure}
The refined structural parameters and the full modes details for both candidates are given in the supplemental material. For convenience, short-lists of the distortive modes for these and other structures used in further discussion are summarized in Table \ref{table:t1}. As expected, the largest amplitudes are assigned to the displacive modes associated with the $\Sigma_2$ ${\bm k}=(1/4,0,-1/4)$ and $R_4^+$ ${\bm k}=(1/2,1/2,1/2)$ order parameters which are identical in both $Pnma$ structures. These order parameters represent antiferroelectric displacements of Bi$^{3+}$ and oxygen along the orthorhombic $a$-axis (pseudocubic $[101]_p$) and antiphase octahedral tilting about this axis, respectively. A schematic representation of the two possible antipolar structures is shown in Fig. \ref{Fig:F2}. The rotations of the octahedra are shown as axial vectors on the Fe/Sc positions (see Ref.\cite{ref:34} for details of this description). The difference between the structures relates to the $\Lambda_3$ ${\bm k}=(-1/4,1/4,1/4)$ and $T_4$ ${\bm k}=(1/2,1/4,1/2)$ order parameters which take different directions in the representation spaces (Table \ref{table:t1}). The conjugated primary distortions involve displacements of Bi$^{3+}$ and unusual octahedral tilting about the $b$-axis (pseudocubic $[010]_p$). The latter implies the $"++--"$ sequence for the corresponding axial vectors (Fig. \ref{Fig:F2}) and has been observed before in some other perovskite systems such as NaNbO$_3$,\cite{ref:35} Bi$_{1-x}$Ln$_x$FeO$_3$ (Ln=lanthanide)\cite{ref:12,ref:17,ref:18,ref:19} and BiFe$_{1-x}$Mn$_x$O$_{3}$.\cite{ref:36}\\
\begin{table*}[t]
\caption{Decomposition of the antipolar and polar structural modifications of BiFe$_{0.5}$Sc$_{0.5}$O$_3$ in respect of the symmetrized displacive modes of the parent cubic $Pm\bar{3}m$ perovskite structure (Bi $1b(1/2,1/2,1/2)$, Fe/Sc $1a(0,0,0)$ and O $3d(1/2,0,0)$). The column "Irrep (${\bm k}$)" shows the irreducible representation of the $Pm\bar{3}m$ space group and the arms of the wave vector star involved (for the propagation vectors inside the Brillouin zone, ${\bm -k}$ are not displayed). The column "Order parameter" lists the projections of the reducible order parameter onto the corresponding irreducible subspace (same symbol in different positions indicates equal order parameter components). The columns "Bi, Fe/Sc, O (site irrep)" display amplitudes of the displacive modes and the corresponding point-group symmetry irreps of the local Wyckoff position (in brackets).}
\centering 
\begin{tabular*}{0.98 \textwidth}{@{\extracolsep{\fill}} l l l l l} 
\hline\hline \\ [-1.5ex]
Irrep (${\bm k}$) & Order parameter & Bi (site irrep) &  Fe/Sc(site irrep) & O (site irrep) \\ [1.0ex] 
\hline \\  [-1.5ex]
   & $Pnma$ origin at (${\bm a_p}/2+{\bm b_p}/2)$ &  &  &  \\ [1.0ex]
\hline \\  [-1.5ex]
$\Sigma_2 (1/4,0,-1/4)$ & $(0,0,0,0,0,0,a,-a,0,0,0,0)$ & $1.15460 (T_{1u})$ & $0.22728 (T_{1u})$ & $-0.13661 (A_{2u})$ \\
& & & & $1.42231 (E_u)$\\
& & & & $-1.19892 (E_u)$\\ [1.5ex]

$R_4^+ (1/2,1/2,1/2)$ & $(\eta ,\eta ,0)$ & & & $-2.61963 (E_u)$\\[1.5ex]

$R_5^+ (1/2,1/2,1/2)$ & $(\delta,-\delta,0)$ & $0.21933 (T1u)$ & &$0.01607 (E_u)$\\[1.5ex]

$\Lambda_3 (-1/4,1/4,1/4)$ & $(0,0,0,0,a,-a,-\sqrt{3}a,-\sqrt{3}a,0,0,0,0,\sqrt{3}a,\sqrt{3}a,a,-a)$ & $-0.60405 (T_{1u})$ & $-0.30443 (T_{1u})$ & $-0.02784 (A_{2u})$\\
$(1/4,1/4,-1/4)$ & & & & $-0.37502 (Eu)$\\
& & & & $-0.26900 (E_u)$\\ [1.5ex]

$T_4 (1/2,1/4,1/2)$ & $(0,0,a,-a,0,0)$ & & & $-0.82390 (E_u)$\\[1.0ex]
\hline \\  [-1.5ex]%
 & $Pnma$ origin at (${\bm a_p}/2+3{\bm b_p}/2)$) & & & \\ [0.5ex]
\hline \\  [-1.5ex]
$\Sigma_2 (1/4,0,-1/4)$ & $(0,0,0,0,0,0,a,-a,0,0,0,0)$ & $1.16028 (T_{1u})$ & $0.23865 (T_{1u})$ & $-0.13661 (A_{2u})$ \\
& & & & $1.42232 (E_u)$\\
& & & & $-1.20460 (E_u)$\\ [1.5ex]

$R_4^+ (1/2,1/2,1/2)$ & $(\eta ,\eta ,0)$ & & & $-2.66455 (E_u)$\\[1.5ex]

$R_5^+ (1/2,1/2,1/2)$ & $(\delta,-\delta,0)$ & $0.20342 (T1u)$ & &$-0.00804 (E_u)$\\[1.5ex]

$\Lambda_3 (-1/4,1/4,1/4)$ & $(0,0,0,0,a,a,\frac{1}{\sqrt{3}}a,-\frac{1}{\sqrt{3}}a,0,0,0,0,-\frac{1}{\sqrt{3}}a,\frac{1}{\sqrt{3}}3a,a,a)$ & $-0.59292 (T_{1u})$ & $-0.15747 (T_{1u})$ & $-0.13454 (A_{2u})$\\
$(1/4,1/4,-1/4)$ & & & & $-0.47161 (Eu)$\\
& & & & $0.44944 (E_u)$\\ [1.5ex]

$T_4 (1/2,1/4,1/2)$ & $(0,0,a,a,0,0)$ & & & $-0.76140 (E_u)$\\[1.0ex]
\hline \\  [-1.5ex]%
& $R3c$ & & & \\[0.5ex]
\hline \\  [-1.5ex]%
$\Gamma_4^- (0,0,0)$ & $(\rho,\rho,\rho)$ & $0.55292 (T_{1u})$ & & $-0.01617 (A_{2u})$ \\
& & & & $-0.77528 (E_u)$ \\ [1.5ex]
$R_4^+ (1/2,1/2,1/2)$ & $(\eta,\eta,\eta)$ & & & $-0.72492 (E_u)$\\ [1.0ex]
\hline \\  [-1.5ex]%
& $Ima2$ & & & \\[1.0ex]
\hline \\  [-1.5ex]%
$\Gamma_4^- (0,0,0)$ & $(\rho,\rho,0)$ & $0.59767 (T_{1u})$ &  & $-0.13973 (A_{2u})$\\
& & & & $-0.69504 (E_u)$ \\  [1.5ex]%
$\Gamma_5^- (0,0,0)$ & $(0,a,-a)$ & & & $-0.04740 (E_u)$\\ [1.5ex]%
$R_4^+ (1/2,1/2,1/2)$ & $(0,\eta ,\eta )$ & & & $-0.95318 (E_u)$\\ [1.5ex]%
$R_5^+ (1/2,1/2,1/2)$ & $(0,\delta,-\delta)$ & $0.16234 (T_{1u})$ & & $-0.13649 (Eu)$\\ [1.0ex]
\hline\hline 
\end{tabular*}
\label{table:t1} 
\end{table*}
\indent The low-temperature neutron diffraction data revealed no structural changes down to 1.5 K. Below T$_N$=220 K, however, an additional scattering to some nuclear reflections at low momentum transfer appeared indicating onset of magnetic ordering with the ${\bm k}=0$ propagation vector (Fig. \ref{Fig:F3}a,b). The magnetic reflections are resolution-limited pointing to a long-range nature of the ordering. This behaviour is consistent with the magnetization data which evidenced a weak ferromagnetic component below T$_N$ (Fig. \ref{Fig:F4}b). The value of the spontaneous moment estimated at 5 K from the magnetization loop (Fig. \ref{Fig:F4}c) was found to be $\sim$0.16 emu/g ($\sim$0.01 $\mu_B$ per Fe/Sc site). Assuming irreducible nature of the magnetic order parameter referring to the paramagnetic $Pnma1'$ space group, the appropriate symmetry restrictions were included in the refinement procedure, which allowed us to reduce significantly the number of possible magnetic structures. The best agreement ($R_{mag}=3.81\%$) with the experimental data was obtained in the antiferromagnetic model with the $m\Gamma_2^+$ symmetry of the magnetic order parameter resulting in the $Pn'm'a$ magnetic space group (Fig. \ref{Fig:F3}c). We followed the notations of ISODISTORT\cite{ref:32} and used "$m$"-letter to distinguish the time-odd representations of paramagnetic gray groups. The suggested model implies the so called $G_y$-type of the spin arrangement where the nearest neighbours have opposite directions of the $y$-spin components (Fig. \ref{Fig:F4}a). 
\begin{figure}[t]
\includegraphics[scale=1.35]{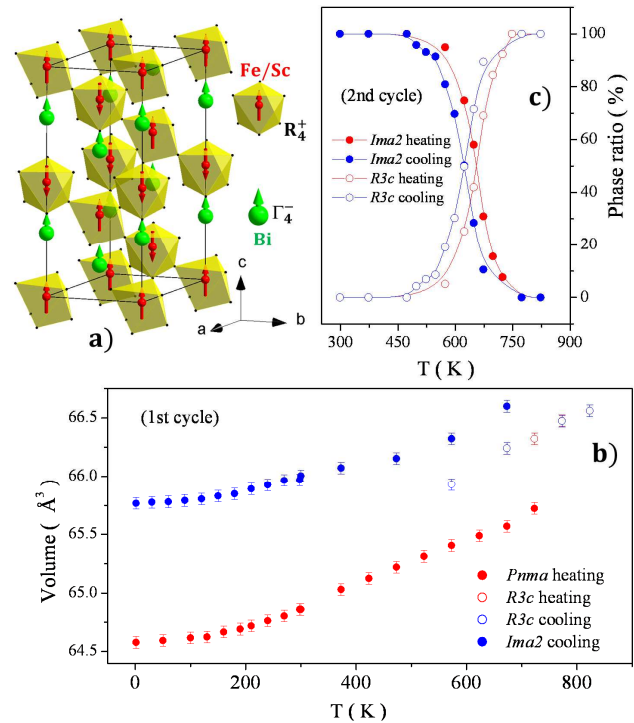}
\caption{(Color online) (a) Polyhedral representation of the high temperature $R3c$ polar structure of BiFe$_{0.5}$Sc$_{0.5}$O$_3$. Arrows, in the Bi and Fe/Sc positions, represent polar (displacements) and axial (octahedral rotations) vectors, respectively. (b) Pseudocubic unit cell volume as a function of temperature in the first thermal heating/cooling cycle ($Pnma$ - a low-temperature phase in the as-prepared sample, $R3c$ - a high-temperature phase, $Ima2$ - a new low-temperature orthorhombic phase). (c) Phase ratio between $R3c$ and $Ima2$ phases as a function of temperature in the second thermal heating/cooling cycle.}
\label{Fig:F5}
\end{figure}
The value of the magnetic moments refined from the data collected at 1.5 K was 2.10(6) $\mu_B$ per Fe/Sc site. The $m\Gamma_2^+$ representation is one-dimensional and appears three times in the decomposition of the reducible magnetic representation on the $8d$ Wyckoff position occupied by Fe ions. One of these two additional orthogonal modes allowed by the $Pn'm'a$ symmetry is ferromagnetic $F_z$ with the moments directed along the $c$-axis (Fig. \ref{Fig:F4}a). This naturally explains the weak ferromagnetic component observed in the magnetization data since both $G_y$ and $F_z$ spin configurations have identical symmetry in the $Pnma$ structure and therefore are expected to be linearly coupled via antisymmetric Dzyaloshinskii-Moria exchange. A direct observation of the $F_z$ component is beyond sensitivity of the powder neutron diffraction experiment. It should be pointed out, however, that the primary $G_y$ mode implies antiferromagnetic spin configuration within a single $8d$ position (Fig. \ref{Fig:F4}a) excluding any possibility of ferrimagnetism.\\
\subsection{High-temperature diffraction study}
To explore the high-temperature structural behaviour of BiFe$_{0.5}$Sc$_{0.5}$O$_3$, X-ray diffraction data were collected in the temperature range between 300 K and 820 K. It was observed that the antipolar orthorhombic $Pnma$ structure is stable up to $\sim$720 K. Slightly below this temperature, a new perovskite phase appears and coexists with $Pnma$. This high-temperature phase was found to be similar to that of BiFeO$_3$ (Fig. \ref{Fig:F5}a) and was successfully refined in the same $R3c$ space group. The refined structural parameters and result of the mode decomposition are summarized in the supplemental material and Table \ref{table:t1}, respectively. The structure involves two primary distortions, $\Gamma_4^-(\rho,\rho,\rho)$-ferroelectric (polar) displacement of Bi$^{3+}$ and oxygen along the pseudocubic $[111]_p$ direction (hexagonal $c$-axis) and $R_4^+(\eta,\eta,\eta)$-antiphase octahedral tilting about this axis (Fig. \ref{Fig:F5}a). The subsequent cooling revealed a remarkable behaviour of BiFe$_{0.5}$Sc$_{0.5}$O$_3$: the structural transition with the phase coexistence between about 670 K and 570 K resulted in a new orthorhombic phase. This phase has a significantly larger unit cell volume (Fig. \ref{Fig:F5}b) in comparison with that of the $Pnma$ structure. The second and all the next subsequent heating/cooling thermal cycles with maximal temperature not exceeding 820 K demonstrated reversible crossover between the new orthorhombic and the polar $R3c$ phases (Fig. \ref{Fig:F5}c). Annealing of the sample above 970 K resulted in its decomposition (see above the experimental part). Thus, the antipolar $Pnma$ structure can only be stabilized at ambient conditions by cooling the system through the high temperature structural phase transition under high pressure. Such behaviour might suggest the existence of the pressure-induced irreversible transition from the new orthorhombic phase to the $Pnma$ one at room temperature.\\ 
\begin{figure}[t]
\includegraphics[scale=1.0]{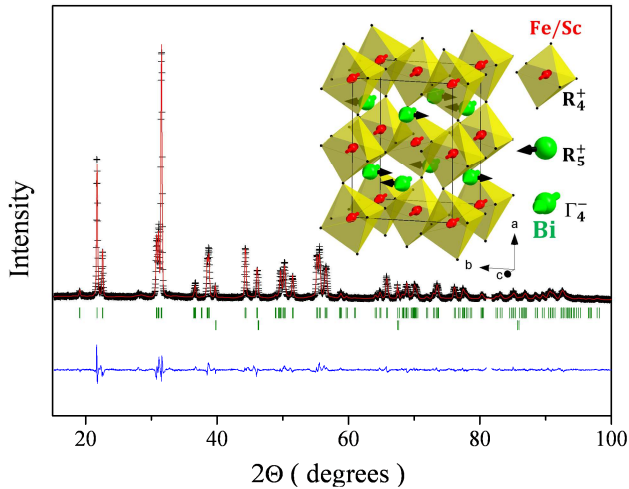}
\caption{(Color online) Rietveld refinement of X-ray diffraction data collected at 300 K on the BiFe$_{0.5}$Sc$_{0.5}$O$_3$ sample annealed at 820 K. The cross symbols and solid (red) line represent the experimental and calculated intensities, respectively, and the line below (blue) is the difference between them. Tick marks (green) indicate the positions of Bragg peaks in the $Ima2$ space group (top) and from the platinum holder (bottom). Inset shows a polyhedral representation of the polar $Ima2$ structure. Arrows, in the Bi and Fe/Sc positions, represent polar (displacements) and axial (octahedral rotations) vectors, respectively.}
\label{Fig:F6}
\end{figure}
\subsection{Crystal and magnetic structures of polar phase}
It was found that both the X-ray and neutron diffraction patterns of the new orthorhombic phase can be indexed using the $2a_p\times \sqrt{2}a_p\times \sqrt{2}a_p$ supercell. The structure is characterized by an extremely distorted metric of the pseudocubic subcell with $a_p<b_p=c_p$ and $\alpha \neq 90^{\circ }$. The only superstructure which was experimentally observed is associated with the $R$-point of symmetry, ${\bm k}=1/2,1/2,1/2$. However, none of the isotropy subgroups related to this propagation vector was able to fit the diffraction data satisfactorily, indicating a reducible nature of the distortions. The solution was therefore searched between the subgroups associated with the coupled order parameters involving $\Gamma $- and $R$-points. The primary candidates were $\Gamma _4^-$ and $R_4^+$ representations of the $Pm\bar{3}m$ space group transforming ferroelectric displacements and antiphase octahedral tilting, respectively. Indeed, the $Ima2$ subgroup which is a result of the intersection between $\Gamma _4^- (\rho,\rho,0)$ and $R_4^+(0,\eta,\eta)$ order parameters provided a good refinement quality for both the X-ray and neutron diffraction data (Fig. \ref{Fig:F6}). Other possibilities consistent with the experimentally observed superstructure and reflection conditions were tested as well and resulted in essentially worse agreement with the experiment. It should be pointed out that taking the $(\rho,\rho,0,\eta,0,0)$ and $(\rho,0,0,\eta,0,\eta)$ directions in the $\Gamma_4^- \oplus  R_4^+$ representation space, one can get two more non-equivalent $Ima2$ subgroups which, however, can be ruled out based on the refinement quality.\\
\begin{table}[t]
\caption{Atomic coordinates and thermal parameters for the annealed BiFe$_{0.5}$Sc$_{0.5}$O$_3$ sample at T=300 K, refined in the $Ima2$ space group with the basis vectors related to the parent cubic cell as: ${\bm a_o}=2{\bm c_p}, {\bm b_o}=-{\bm a_p}+{\bm b_p}, {\bm c_o}=-{\bm a_p}-{\bm b_p}$ and origin at $(0,0,0)$. Unit cell parameters: $a_o=7.8836(2)\AA$, $b_o=5.7538(2)\AA$ and $c_o=5.8075(2)\AA$, $R_{Bragg(X-ray)}=4.91\%$, $R_{Bragg(neutr)}=5.07\%$}
\centering 
\begin{tabular*}{0.48\textwidth}{@{\extracolsep{\fill}} l c c c c c} 
\hline\hline\\  [-1.5ex]
Atom & Site & $x$ & $y$ & $z$ & $B_{iso}$ \\ [1.0ex] 
\hline\\[-1.5ex] 
Bi & 4$b$ & 0.25 & 0.5201(7) & -0.0740(6) & 0.4(1)\\ 
Fe/Sc & 4$a$ & 0 & 0 & 0 & 0.8(1)\\
O1 & 8$c$ & 0.0477(3) & 0.7697(7) & 0.2870(9) & 1.0(1)\\
O2 & 4$b$ & 0.25 & 0.0715(9) & 0.065(1) & 1.0(1)\\ [1.0ex]
\hline\hline  
\end{tabular*}
\label{table:t2} 
\end{table}
\indent The structural parameters of the $Ima2$ phase and the corresponding displacive mode characteristics are summarized in Table \ref{table:t2} and Table \ref{table:t1}, respectively. The structure represents a new polar variant of the perovskite lattice, which combines ferroelectric displacements of Bi$^{3+}$ ions along the $[110]_p$ pseudocubic direction and antiphase tilting of the Fe/ScO$_6$ octahedra about this direction (Fig. \ref{Fig:F6}, inset). Using the notations adopted by Stokes et al.,\cite{ref:37} the distortions can be presented as $a_+^- a_+^- b_0^0$, where the superscripts and subscripts indicate ferroelectric displacements of Bi and octahedral tilting respectively. A remarkable feature is the presence of the $R_5^+(0,\delta,-\delta )$ distortive mode which implies antiferroelectric displacements of Bi$^{3+}$ along the $[1\bar{1}0]_p$ pseudocubic direction. Thus, BiFe$_{0.5}$Sc$_{0.5}$O$_3$ in the polar $Ima2$ phase is actually a canted ferroelectric. Let us discuss this point in more details and demonstrate the symmetry reasons for this phenomenon. The $R_5^+(0,\delta,-\delta)$ distortion can be treated as a secondary order parameter whose presence is allowed due to its identical symmetry with the $R_4^+(0,\eta,\eta)$ one (both order parameters reduce the parent cubic symmetry $Pm\bar{3}m$ down to the orthorhombic $Imma$). The coupling is provided through the general free-energy invariant:
\begin{eqnarray}
\delta_1\eta _1\eta _2^2 - \delta_1\eta _1\eta _3^2 - \delta_2\eta _2\eta _1^2 + \nonumber \\
\delta_2\eta _2\eta _3^2 + \delta_3\eta _3\eta _1^2 - \delta_3\eta _3\eta _2^2
\label{eq:E1}
\end{eqnarray}
\begin{figure}[t]
\includegraphics[scale=1.0]{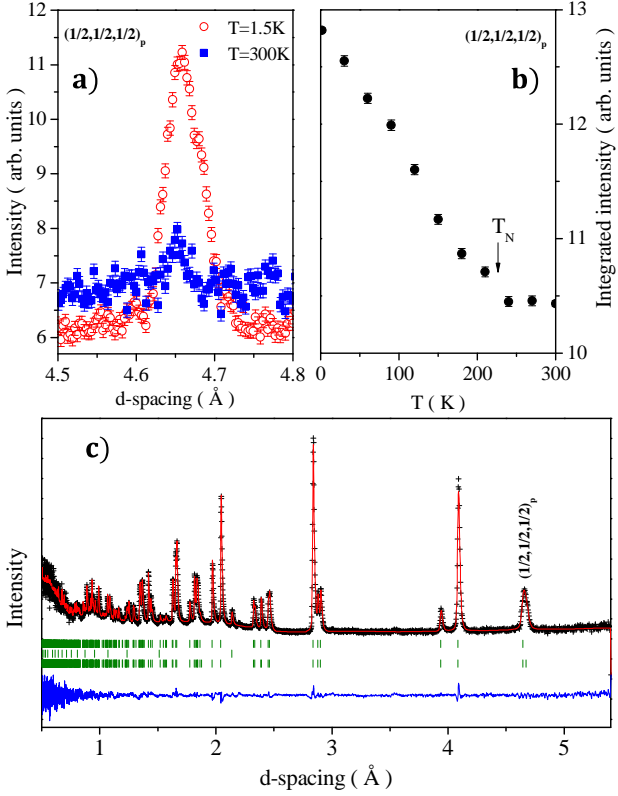}
\caption{(Color online) (a) Neutron diffraction patterns for the annealed BiFe$_{0.5}$Sc$_{0.5}$O$_3$ sample at the vicinity of the strongest $(1/2,1/2,1/2)_p$ magnetic peak collected above and below T$_N$. (b) Integrated intensity of the $(1/2,1/2,1/2)_p$ peak as a function of temperature. (c) Rietveld refinement of the neutron diffraction data collected at 1.5 K. The cross symbols and solid line (red) represent the experimental and calculated intensities, respectively, and the line below (blue) is the difference between them. Tick marks (green) indicate the positions of Bragg peaks: nuclear (top), vanadium can (middle) and magnetic ${\bm k}=0$ (bottom).}
\label{Fig:F7}
\end{figure}
\begin{figure}[t]
\includegraphics[scale=1.0]{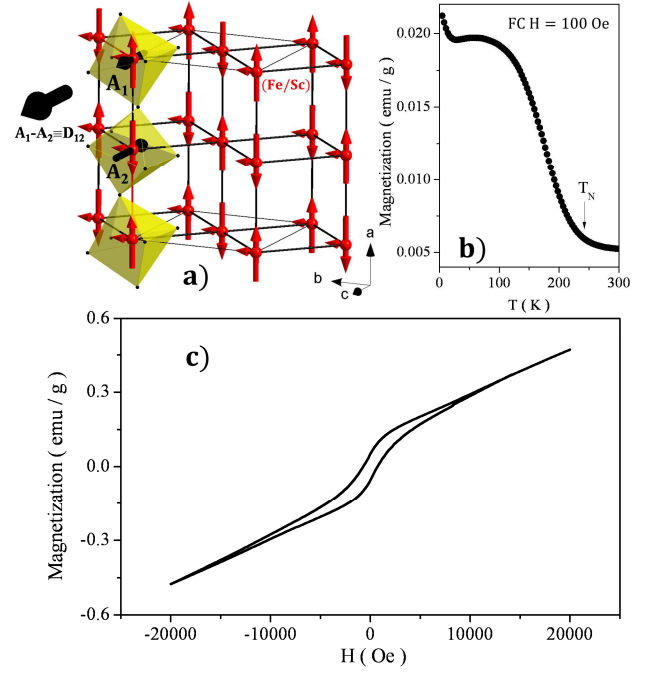}
\caption{(Color online) (a) Magnetic structure of the annealed BiFe$_{0.5}$Sc$_{0.5}$O$_3$ sample with the $Im'a2'$ magnetic space group. The largest spin component $\sim$2.20(5) $\mu_B$ found in the neutron diffraction experiment is along the $a$-axis ($G_x$). The orthogonal ferromagnetic component, along the $b$-axis ($F_y$) allowed by the $Im'a2'$ symmetry, is shown as well. For two Fe/Sc sites, axial vectors ${\bm A_1}$ and ${\bm A_2}$ along the $c$-axis, representing octahedral tilting and contributing to the Dzyaloshinskii vector ${\bm D_{12}}$ are displayed as antiparallel arrows (black). (b) Magnetization as a function of temperature, measured under the magnetic field H=100 Oe after cooling under the same field. (c) Magnetization loop measured at 5 K after cooling in H=20000 Oe.}
\label{Fig:F8}
\end{figure}
where $\delta_i$ and $\eta_i (i=1-3)$ are components of the $R_5^+$ and $R_4^+$ order parameters, respectively. The irrep matrices for the space group generators, used to construct the invariant, are summarized in Table \ref{table:t3}. The fourth-degree power of the invariant is imposed by the translational symmetry, since both representations are associated with the ${\bm k}=1/2,1/2,1/2$ propagation vector. For the equilibrium value of the $R_4^+$ order parameter $(0,\eta_2,\eta_3; \eta_2=\eta_3=\eta)$, the expression (\ref{eq:E1}) reduces down to $\delta_2 \eta^3 - \delta_3 \eta^3$ which requires $(\delta_2=-\delta_3)$, in full agreement with the results of the mode decomposition (Table \ref{table:t1}). It is straightforward to verify that minimization of the relevant part of the free-energy, $\Phi =\alpha \delta^2+\gamma \delta \eta^3$, results in a non-zero equilibrium value for $\delta =-\frac{\gamma}{2\alpha} \eta^3$. Thus, the presence of the octahedral tilting $R_4^+(0,\eta,\eta)$ is sufficient to generate a finite $R_5^+(0,\delta,-\delta)$ antiferroelectric displacement of Bi$^{3+}$. This consideration, therefore, is valid for the antipolar phase as well, where the $R_4^+(\eta,\eta,0)$ instability is a part of the structural distortions in the $Pnma$ space group (Table \ref{table:t1}). The expression (\ref{eq:E1}), however, vanishes for the $(\eta,\eta,\eta)$ order parameter direction taking place in the high-temperature polar $R3c$ phase.
\begin{table*}[t]
\caption{Matrix of the irreducible representations for generators of the parent $Pm\bar{3}m1'$ space group. $T$ is the time reversal operator.}
\centering 
\begin{tabular*}{0.98\textwidth}{@{\extracolsep{\fill}} l c c c c} 
\hline \hline \\ [-1.5ex]
Irrep & $4^+_{(001)}$ & $3^+_{(111)}$ & $\bar{1} $ & $T$ \\ [0.5ex] 
\hline \\ [-1.0ex] 
$\Gamma_4^-$ & $\left ( \begin{array}{ccc} 0 & -1 & 0 \\ 1 & 0 & 0 \\ 0 & 0 & 1 \end{array} \right )$ & $\left ( \begin{array}{ccc} 0 & 0 & 1 \\ 1 & 0 & 0 \\ 0 & 1 & 0 \end{array} \right )$ & $\left ( \begin{array}{ccc} -1 & 0 & 0 \\ 0 & -1 & 0 \\ 0 & 0 & -1 \end{array} \right )$ & $\left ( \begin{array}{ccc} 1 & 0 & 0 \\ 0 & 1 & 0 \\ 0 & 0 & 1 \end{array} \right )$ \\ [4.0ex]

$m\Gamma_4^+$ & $\left ( \begin{array}{ccc} 0 & -1 & 0 \\ 1 & 0 & 0 \\ 0 & 0 & 1 \end{array} \right )$ & $\left ( \begin{array}{ccc} 0 & 0 & 1 \\ 1 & 0 & 0 \\ 0 & 1 & 0 \end{array} \right )$ & $\left ( \begin{array}{ccc} 1 & 0 & 0 \\ 0 & 1 & 0 \\ 0 & 0 & 1 \end{array} \right )$ & $\left ( \begin{array}{ccc} -1 & 0 & 0 \\ 0 & -1 & 0 \\ 0 & 0 & -1 \end{array} \right )$ \\ [4.0ex]

$R_4^+$ & $\left ( \begin{array}{ccc} 1 & 0 & 0 \\ 0 & 0 & -1 \\ 0 & 1 & 0 \end{array} \right )$ & $\left ( \begin{array}{ccc} 0 & 0 & 1 \\ 1 & 0 & 0 \\ 0 & 1 & 0 \end{array} \right )$ & $\left ( \begin{array}{ccc} 1 & 0 & 0 \\ 0 & 1 & 0 \\ 0 & 0 & 1 \end{array} \right )$ & $\left ( \begin{array}{ccc} 1 & 0 & 0 \\ 0 & 1 & 0 \\ 0 & 0 & 1 \end{array} \right )$ \\ [4.0ex]

$mR_4^+$ & $\left ( \begin{array}{ccc} 1 & 0 & 0 \\ 0 & 0 & -1 \\ 0 & 1 & 0 \end{array} \right )$ & $\left ( \begin{array}{ccc} 0 & 0 & 1 \\ 1 & 0 & 0 \\ 0 & 1 & 0 \end{array} \right )$ & $\left ( \begin{array}{ccc} 1 & 0 & 0 \\ 0 & 1 & 0 \\ 0 & 0 & 1 \end{array} \right )$ & $\left ( \begin{array}{ccc} -1 & 0 & 0 \\ 0 & -1 & 0 \\ 0 & 0 & -1 \end{array} \right )$ \\ [4.0ex]

$R_5^+$ & $\left ( \begin{array}{ccc} -1 & 0 & 0 \\ 0 & 0 & 1 \\ 0 & -1 & 0 \end{array} \right )$ & $\left ( \begin{array}{ccc} 0 & 0 & 1 \\ 1 & 0 & 0 \\ 0 & 1 & 0 \end{array} \right )$ & $\left ( \begin{array}{ccc} 1 & 0 & 0 \\ 0 & 1 & 0 \\ 0 & 0 & 1 \end{array} \right )$ & $\left ( \begin{array}{ccc} 1 & 0 & 0 \\ 0 & 1 & 0 \\ 0 & 0 & 1 \end{array} \right )$ \\ [4.0ex]

\hline
\end{tabular*}
\label{table:t3} 
\end{table*}
In addition to the above free-energy invariant, there is another coupling term in the $Ima2$ structure relating the $R_5^+$ antiferroelectric displacement of Bi$^{3+}$ and antiphase octahedral tilting via the polar distortions:
\begin{eqnarray}
\delta_1\eta _1\rho_1^2 - \delta_1\eta _1\rho_2^2 + \delta_2\eta _2\rho _2^2 - \nonumber \\
\delta_2\eta _2\rho _3^2 - \delta_3\eta _3\rho_1^2 + \delta_3\eta _3\rho _3^2
\label{eq:E2}
\end{eqnarray}
where $\rho_i, (i=1-3)$ are components of the polar $\Gamma_4^-$ order parameter, which reduces down to $2\delta \eta \rho^2$ for the equilibrium values of $R_4^+(0,\eta,\eta)$ and $\Gamma_4^-(\rho,\rho,0)$. This coupling term yields $\delta=-\frac{\gamma}{2\alpha}\eta \rho^2$ indicating that switching of the polarization, $\rho $, does not affect the antiferroelectric displacements. Which of the above two free-energy invariants is actually dominating can be concluded only based on a thorough analysis of the critical behaviour of the $R_5^+$, $R_4^+$ and $\Gamma_4^-$ order parameters. However, such an analysis is impossible from the direct diffraction measurements due to the existence of the high temperature $R3c$ phase cancelling both types of coupling.\\
\indent The polar orthorhombic $Ima2$ phase was found from the low-temperature neutron diffraction measurements to be stable down to 1.5 K. Similar to the $Pnma$ modification, the onset of the magnetic ordering in $Ima2$ occurs at T$_N \sim$220 K (Fig. \ref{Fig:F7}a,b). Qualitatively, the magnetic scattering is very similar in both phases; however, the magnetic reflections are slightly broader in the annealed sample, which might indicate a shorter correlation length. In principle, the annealing procedure can promote some local short-range order between Fe and Sc ions undetectable in the diffraction experiments. This can significantly affect the nearest neighbour superexchange interactions, which are expected to be dominant through the $e_g - p_{\sigma}-e_g$ orbital overlapping. An alternative explanation could be the presence of a small incommensurability due to the polar nature of the $Ima2$ structure similar to that observed in the parent compound BiFeO$_3$, with satellites positions too close to the commensurate peak to be resolved with the current experimental resolution. In such a case, the ordering with the ${\bm k}=0$ propagation vector and the $G$-type of the antiferromagnetic spin arrangement is just a first approximation to the magnetic structure. The best fitting quality (Fig. \ref{Fig:F7}c) of the experimental data ($R_{mag}=4.13 \%$) assuming the commensurate variant and irreducible magnetic order parameter was obtained in the $m\Gamma_4$ representation of the $Ima21'$ space group, implying the spins $\sim$2.20(5) $\mu_B$ per Fe/Sc site are aligned along the $a$-axis ($G_x$-mode, see Fig. \ref{Fig:F8}a). The refinement can be slightly improved by some admixture of the $m\Gamma_1$ irrep ($G_z$-mode), which however can be a consequence of the peaks broadening. In the incommensurate scenario, both representations can be combined into the Lifshitz-type free-energy invariant $\zeta  \frac{\partial \vartheta  }{\partial x} - \vartheta \frac{\partial \zeta}{\partial x}$ (where $\zeta $ and $\vartheta $ are the magnetic order parameters transformed by the $m\Gamma_4$ and $m\Gamma_1$ representations, respectively) with the gradient terms favouring the inhomogeneous state. This can happen only if the magnetocrystalline anisotropy, which splits the $G_x$, $G_y$ and $G_z$ components of the exchange multiplet, is vanishingly small.\\
\indent In the commensurate case, the $m\Gamma _4$ representation appears twice in the decomposition of the reducible magnetic representation on the $4a$ Wyckoff position of Fe/Sc, resulting in the $Im'a2'$ magnetic space group.\cite{ref:38} The latter allows an admixture of a weak ferromagnetic component along the $b$-axis (Fig. \ref{Fig:F8}a). Thus, magnetization measurements can discriminate between commensurate and incommensurate scenario, since only the former allows a non-zero spontaneous magnetization. Indeed, a weak spontaneous moment $\sim 0.13$ emu/g was found in the measurements of the magnetization as a function of field at 5 K (Fig. \ref{Fig:F8}c), revealing the weak-ferromagnetic nature of the $Ima2$ phase. The ferromagnetic component develops below T$_N\sim$ 220 K (Fig. \ref{Fig:F8}b) in agreement with the neutron diffraction data. Thus, BiFe$_{0.5}$Sc$_{0.5}$O$_3$ in the $Ima2$ structural modification is a rare example of a polar weak-ferromagnet. This fact makes BiFe$_{1-x}$Sc$_x$O$_3$ a very attractive muliferroic system, since the magnetic transition can be easily tuned above room temperature by slight increasing the Fe content ($x \sim 0.42$ assuming a linear dependence of T$_N$ on $x$).\\
\section{Discussion}
As it was shown above, the metastable perovskite BiFe$_{0.5}$Sc$_{0.5}$O$_3$ exhibits an irreversible high-temperature phase transition which provides a way to get this material in two different polymorphs, namely, antipolar with the $Pnma$ space group and polar with the $Ima2$ symmetry. This remarkable behaviour indicates that the high temperature transition occurs between $R3c$ and $Pnma$ at 6 GPa and between $R3c$ and $Ima2$ phases at ambient pressure. This points to the possibility of irreversible pressure-induced transition from the polar to antipolar modification ($Ima2 \longrightarrow Pnma$) even at room temperature. The existence of the two polymorphs provides a good playground to study a structure-properties relationship. In this respect, multiferroic properties are of a particular interest since both modifications involve dipole, magnetic and elastic order parameters. The $Ima2$ phase is a rare example where ferroelectrisity coexists with a ferromagnetic component, providing an intriguing possibility to explore the coupling between them. Let us discuss the interference between the different order parameters in a perovkite structure based on a phenomenological approach. The most general consideration is based on the parent $Pm\bar{3}m1'$ cubic symmetry and is able to draw conclusions relevant for all structural phases described in the previous sections as well as for any other perovskites. The ferromagnetic and the $G$-type antiferromagnetic spin configurations transform as the time-odd $m\Gamma_4^+ ({\bm k}=0)$ and $mR_4^+ ({\bm k}=1/2,1/2,1/2)$ irreducible representations, respectively. The translation and time-reversal symmetries imply that the lowest degree free-energy coupling invariant should be trilinear with a time-even physical quantity which breaks the translational invariance in the same way as the antiferromagnetic order does. The relevant analysis drew us to the conclusion that the appropriate quantity is the $R_4^+$ axial distortion associated with octahedral tilting, which is the most common structural distortion in the perovskite family. The relevant free-energy term can be represented as:
\begin{eqnarray}
\eta_1 \mu_1 \xi_3 - \eta_1 \mu_2 \xi_2 + \eta_2 \mu_2 \xi_1 -    \nonumber \\
\eta_2 \mu_3 \xi_3 - \eta_3 \mu_1 \xi_1 + \eta_3 \mu_3 \xi_2
\label{eq:E3}
\end{eqnarray}
where $\eta_i$, $\mu_i$ and $\xi_i (i=1-3)$ are components of the $R_4^+$, $m\Gamma_4^+$ and $mR_4^+$ order parameters, respectively. Using the projection operator technique, we can symmetrize the atomic spin components $S^j (j=x,y,z)$ and the axial vectors $A^j (j=x,y,z)$ representing rotations of the ochahedra, to express the order parameters in terms of the localized atomic basis functions. It gives us the advantage to write down the free-energy invariant in terms of the spin components. The interactions between nearest neighbours are dominant in the system as follows from the experimentally determined magnetic structures (see Figs. \ref{Fig:F4}a and \ref{Fig:F8}a). Therefore, it is enough to consider only two neighbour spins: $\#1$ at the coordinate origin and $\#2$ separated by any of the shortest lattice translations $\pm [100]_p$, $\pm [010]_p$ or $\pm [001]_p$ with the localized pseudovector functions related through the appropriate Fourier transform.
\begin{eqnarray}
\eta_1 = A_1^z - A_2^z, \eta_2 = A_1^x - A_2^x, \eta_3 = A_1^y - A_2^y  \nonumber \\
\mu_1 = S_1^x + S_2^x, \mu_2 = S_1^y + S_2^y, \mu_3 = S_1^z + S_2^z \nonumber \\
\xi_1 = S_1^z - S_2^z, \xi_2 = S_1^x - S_2^x, \xi_3 = S_1^y - S_2^y
\label{eq:E4}
\end{eqnarray}
Neglecting the terms like $S_iS_k (i=k)$ which represent single ion effects, we can express the coupling invariant (\ref{eq:E3}) as:
\begin{eqnarray}
2(A_1^z-A_2^z)[S_1^y S_2^x - S_1^x S_2^y] + 2(A_1^x-A_2^x) [S_1^y S_2^z - \nonumber \\
S_1^z S_2^y] + 2(A_1^y - A_2^y)[S_1^x S_2^z - S_1^z S_2^x]
\label{eq:E5}
\end{eqnarray}
This expression is identical to the well-known phenomenological term ${\bm D_{12}}[{\bm S_1}\times {\bm S_2}]$ for the Dzyaloshinskii-Moria antisymmetric exchange, where the Dzyaloshinskii vector, ${\bm D_{12}}$, is represented by the antiferroaxial vector ${\bm A_{12}={\bm A_1} - {\bm A_2} ({\bm A_{12}\equiv {\bm D_{12}}})}$. This vector is analogous to an antiferromagnetic vector and reflects the antiferrodistortive character of the octahedral tilting (Fig. \ref{Fig:F8}a). Thus, the obtained result indicates that the octahedral tilting in perovskites with a $G$-type antiferromagnetic order is the main distortion responsible for the weak-ferromagnetism and the antiferroaxial vector ${\bm A_{12}}$ represents the relevant part of the Dzyaloshinskii vector. This general conclusion is fully consistent with the recent result by Zvezdin and Pyatakov et al.,\cite{ref:39} obtained for BiFeO$_3$ based on the Keffer formula. The octahedral tilting induces a spin canting when the antiferromagnetically ordered spins are confined in the plane perpendicular to the tilting axis (Fig. \ref{Fig:F8}a). In the case of BiFe$_{0.5}$Sc$_{0.5}$O$_3$, both $Pnma$ and $Ima2$ phases involve octahedral tilting resulting in the experimentally observed weak-ferromagnetic component. In the $Pn'm'a$ structure, the spins are along the $b$-axis (psedocubic $[010]_p$) in the $G_y$-mode and the octahedra are tilted about the $a$-axis (pseudocubic $[101]_p$) generating the spin canting along the $c$-axis (pseudocubic $[\bar{1} 01]_p$). Similarly, in the $Im'a2'$ structure, the octahedral tilting occurring about the $c$-axis (pceudocubic $[110]_p$) couples the weak-ferromagnetic component along the $b$-axis (pseudocubic $[1\bar{1}0]_p$) to the antiferromagnetically arranged spins along the $a$-axis (psedocubic $[001]_p$).\\
\indent We do not consider the part of the antisymmetric exchange favouring inhomogeneus state (like the cycloidal modulation in BiFeO$_3$) since it has not been observed in BiFe$_{0.5}$Sc$_{0.5}$O$_3$ perovskite, in spite of the fact that the necessary symmetry conditions are fulfilled. We just note that the relevant part of the Dzyaloshinskii vector responsible for this mechanism is associated with the polar displacements of oxygen ($\Gamma_4^-$ displacive modes of oxygen in Table \ref{table:t1}). The reason why this mechanism dos not induce inhomogeneous state in BiFe$_{0.5}$Sc$_{0.5}$O$_3$ is probably the high concentration of the non-magnetic Sc ions which work as local defects destroying the coherence of the system, necessary for realization of the long-period modulated structure.\\
\indent Let us discuss the coupling between polarization and ferromagnetic order parameter transforming as $\Gamma_4^-$ and $m\Gamma_4^+$ irreducible representations of the $Pm\bar{3}m1'$ space group, respectively. We denote the corresponding components of the macroscopic quantities as $(P_x,P_y,P_z)$ and $(M_x,M_y,M_z)$. The lowest degree free-energy invariant which can be constructed using the matrices from Table \ref{table:t3} is the following:
\begin{eqnarray}
 P_x P_y M_x M_y + P_x P_z M_x M_z+P_y P_z M_y M_z 
\label{eq:E6}
\end{eqnarray}
We can use the fact that the weak-ferromagnetic components, generated by the octahedral tilting and contributing to the macroscopic magnetization, are secondary order parameters whose equilibrium values can be found by minimization of the relevant free-energy part truncated at the lowest degree coupling term (\ref{eq:E3}):
\begin{eqnarray}
\mu _1 \sim (\eta_1 \xi_3 - \eta_3 \xi_1)  \nonumber \\
\mu _2 \sim (\eta_2 \xi_1 - \eta_1 \xi_2) \nonumber \\
\mu _3 \sim (\eta_3 \xi_2 - \eta_2 \xi_3)
\label{eq:E7}
\end{eqnarray}
The phenomenological coefficients are not essential for our consideration and can be skipped. In the case of the polar $Ima2$ phase, $P_z=0$ and the expression (\ref{eq:E6}) is reduced down to $P_xP_yM_xM_y$. After substitution of the expression (\ref{eq:E7}) for $\mu_2$ instead of $M_y$ (both quantities have identical transformational properties) and taking into account the equilibrium values for the $R_4^+$ and $mR_4^+$ order parameters, we get the fifth degree coupling term $M_xP_xP_y\eta_2\xi_1$. The term indicates that $M_x \sim P_xP_y\eta_2\xi_1$ and analogously, $M_y \sim -P_xP_y\eta_3\xi_1$ and describes the magnetoelectric contribution (generated by the spontaneous polarization\cite{ref:7}) to the weak-ferromagnetism. One can see that at least two non-zero components of the polarization are necessary to activate this mechanism; one component breaks the inversion and creates conditions for linear magnetoelectric effect, another component works as an internal electric filed, generating magnetization. The higher order of these free-energy terms, however, implies that the magnetoelectric contribution is small in comparison with the mechanism caused by the octahedral tilting. This conclusion is also relevant for the parent BiFeO$_3$ compound. It should be pointed out that the presence of the octahedral tilting is also an essential ingredient to activate the magnetoelectric mechanism.\\
\indent The obtained above relations written in the form $M_x \sim E_yP_x\eta_2\xi_1$ and $M_y \sim -E_xP_y\eta_3\xi_1$, where $E_x$ and $E_y$ are the components of electric field along the $a_p$- and $b_p$-axes, respectively and the spontaneous polarization is considered as a frozen distortion, describe also the linear magnetoelectric properties of the $Im'a2'$ phase. For instance, application of the electric field ${\bm E}$, along the $[\bar{1}10]_p$-direction (orthorhombic $b$-axis) implies a generation of the ferromagnetic component along the $[110]_p$-axis (orthorombic $c$-axis) which corresponds to a non-zero $\alpha_{23}$ component of the magnetoelectric tensor (referring to the orthorhombic setting) in agreement with the requirement of the $m'm2'$ magnetic point group controlling the macroscopic properties of the $Im'a2'$ phase.\\
\section{Conclusion}
\indent A metastable perovskite phase in the system BiFe$_{0.5}$Sc$_{0.5}$O$_3$ is stabilized under high pressure and high temperature conditions. The samples quenched under high pressure have orthorhombic $Pnma$ symmetry with the $\sqrt{2}a_p\times 4a_p\times 2\sqrt{2}a_p$ superstructure resulted from antiferroelectric displacements of Bi$^{3+}$, quadrupling the pseudocubic perovskite subcell in all three directions, and associated $++--$ octahedral tilting. Combining these distortion types, two non-equivalent structures with the $Pnma$ space group can be obtained. The antipolar phase transforms into the polar $R3c$ one on heating. The transformation takes place in the temperature range of 700 - 720 K, where both phases coexist. The rhombohedral $R3c$ phase is isostructural to the polar phase of undoped BiFeO$_3$ and characterized by ferroelectric displacements of Bi$^{3+}$ along $[111]_p$ and octahedral tilting about this direction. The $Pnma\longrightarrow R3c$ transition is irreversible at ambient pressure and subsequent cooling below 670 K results in appearance of an orthorhombic phase with the $Ima2$ symmetry and the $2a_p\times \sqrt{2}a_p\times \sqrt{2}a_p$ supercell. This orthorhombic modification of BiFe$_{0.5}$Sc$_{0.5}$O$_3$ is a new type of a polar perovskite structure, where the ferroelectric displacements of Bi$^{3+}$ cations along the $[110]_p$ pseudocubic direction are combined with the antiphase octahedral tilting about the polar axis. The primary distortions couple antiferroelectric displacements of Bi$^{3+}$ along the $[1\bar{1}0]_p$ direction as well, resulting in a canted ferroelectricity. Both the polar $Ima2$ and antipolar $Pnma$ polymorphs exhibit a long-range $G$-type of antiferromagnetic order with a weak-ferromagnetic component below T$_N\sim$ 220 K. The high transition temperature indicates a great potentiality of the system to exhibit the room temperature multiferroic properties by a small modification of the Fe/Sc ratio. The weak-ferromagnetism in both phases is caused by the presence of the antiphase octahedral tilting whose axial nature directly represents the relevant part of Dzyaloshinskii vector. The magnetoelectric contribution to the spontaneous ferromagnetic moment allowed in the polar $Ima2$ phase is described by a fifth-degree free-energy invariant and is expected to be small in comparison with the value generated by the tilting of oxygen octahedra.\\

\begin{table*}[h] 
\caption{({\bf Supplemental material}) Atomic coordinates and thermal parameters for the as-prepared BiFe$_{0.5}$Sc$_{0.5}$O$_3$ sample at T=300 K refined in the $Pnma$ space group with the basis vectors related to the parent cubic cell as ${\bm a_o}={\bm a_p}+{\bm c_p}$, ${\bm b_o}=4{\bm b_p}$, ${\bm c_o}=-2{\bm a_p}+2{\bm c_p}$ and origin at ${\bm a_p}/2+{\bm b_p}/2$. Unit cell parameters: $a_o=5.6950(3)\AA$, $b_o=15.9126(5)\AA$ and $c_o=11.4512(5)\AA$, $R_{Bragg(X-ray)}=5.55\%$, $R_{Bragg(neutr)}=6.83\%$}
\centering 
\begin{tabular*}{0.98\textwidth}{@{\extracolsep{\fill}} l l l l l l} 
\hline\hline\\  [-1.5ex]
Atom & Site & $x$ & $y$ & $z$ & $B_{iso}$ \\ [1.0ex] 
\hline\\[-1.5ex] 
Bi1 & 8$d$ & 0.3101(8) & 0.0105(3) & 0.1330(4) & 0.6(1)\\ 
Bi2 & 4$c$ & 0.301(2) & 0.25 & 0.1118(6) & 0.6(1)\\ 
Bi3 & 4$c$ & 0.218(2) & 0.25 & 0.6349(6) & 0.6(1)\\ 
Fe/Sc1 & 8$d$ & 0.764(2) & 0.8682(8) & 0.122(1) & 0.14(9)\\
Fe/Sc1 & 8$d$ & 0.756(3) & 0.3783(9) & 0.125(1) & 0.14(9)\\
O1 & 8$d$ & -0.041(3) & 0.872(1) & -0.014(2) & 0.5(1)\\
O2 & 8$d$ & -0.001(5) & 0.844(1) & 0.501(3) & 0.5(1)\\
O3 & 8$d$ & 0.567(4) & 0.898(1) & 0.780(2) & 0.5(1)\\ 
O4 & 8$d$ & 0.530(4) & 0.399(1) & 0.760(2) & 0.5(1)\\
O5 & 8$d$ & 0.698(3) & -0.002(2) & 0.076(2) & 0.5(1)\\
O6 & 4$c$ & 0.718(5) & 0.25 & 0.157(3) & 0.5(1)\\
O7 & 4$c$ & 0.825(6) & 0.25 & 0.659(3) & 0.5(1)\\ [1.0ex]
\hline\hline  
\end{tabular*}
\label{table:t4} 
\end{table*}

\begin{table*}[h] 
\caption{({\bf Supplemental material}) Atomic coordinates and thermal parameters for the as-prepared BiFe$_{0.5}$Sc$_{0.5}$O$_3$ sample at T=300 K refined in the $Pnma$ space group with the basis vectors related to the parent cubic cell as ${\bm a_o}={\bm a_p}+{\bm c_p}$, ${\bm b_o}=4{\bm b_p}$, ${\bm c_o}=-2{\bm a_p}+2{\bm c_p}$ and origin at ${\bm a_p}/2+3{\bm b_p}/2$. Unit cell parameters: $a_o=5.6953(3)\AA$, $b_o=15.9124(5)\AA$ and $c_o=11.4511(5)\AA$, $R_{Bragg(X-ray)}=5.40\%$, $R_{Bragg(neutr)}=6.83\%$}
\centering 
\begin{tabular*}{0.98\textwidth}{@{\extracolsep{\fill}} l l l l l l} 
\hline\hline\\  [-1.5ex]
Atom & Site & $x$ & $y$ & $z$ & $B_{iso}$ \\ [1.0ex] 
\hline\\[-1.5ex] 
Bi1 & 4$c$ & 0.181(2) & 0.25 & 0.6219(6) & 0.5(1)\\ 
Bi2 & 4$c$ & 0.299(2) & 0.25 & 0.1434(6) & 0.5(1)\\ 
Bi3 & 8$d$ & 0.2931(8) & -0.0106(2) & 0.1237(4) & 0.5(1)\\ 
Fe/Sc1 & 8$d$ & 0.762(2) & 0.620(1) & 0.122(2) & 0.10(9)\\
Fe/Sc1 & 8$d$ & 0.759(2) & 0.128(1) & 0.126(2) & 0.10(9)\\
O1 & 8$d$ & 0.038(3) & 0.620(1) & 0.013(2) & 0.4(1)\\
O2 & 8$d$ & 0.021(4) & 0.594(1) & 0.506(3) & 0.4(1)\\
O3 & 8$d$ & 0.530(4) & 0.655(1) & 0.770(2) & 0.4(1)\\ 
O4 & 8$8$ & 0.567(4) & 0.142(1) & 0.770(2) & 0.4(1)\\
O5 & 4$c$ & 0.820(6) & 0.25 & 0.591(3) & 0.4(1)\\
O6 & 4$c$ & 0.710(5) & 0.25 & 0.062(2) & 0.4(1)\\
O7 & 8$d$ & 0.699(3) & 0.004(2) & 0.159(2) & 0.4(1)\\ [1.0ex]
\hline\hline  
\end{tabular*}
\label{table:t5} 
\end{table*}

\begin{table*}[t]
\caption{ ({\bf Supplemental material}) Atomic coordinates and thermal parameters for BiFe$_{0.5}$Sc$_{0.5}$O$_3$ at T=820 K refined in the $R3c$ space group with the basis vectors related to the parent cubic cell as ${\bm a_h}=-{\bm a_p}+{\bm b_p}, {\bm b_h}=-{\bm b_p}+{\bm c_p}, {\bm c_h}=2{\bm a_p}+2{\bm b_p}+2{\bm c_p}$ and origin at (0,0,0). Unit cell parameters: $a_h=5.6910(1)\AA$, and $c_h=14.2070(4)\AA$, $R_{Bragg(X-ray)}=4.43\%$.}
\centering 
\begin{tabular*}{0.98\textwidth}{@{\extracolsep{\fill}} l c c c c c} 
\hline\hline\\  [-1.5ex]
Atom & Site & $x$ & $y$ & $z$ & $B_{iso}$ \\ [1.0ex] 
\hline\\[-1.5ex] 
Bi & 6$a$ & 0 & 0 & 0.2779(4) & 1.5(1)\\ 
Fe/Sc & 6$a$ & 0 & 0 & 0 & 1.5(1)\\
O & 18$b$ & 0.631(2) & 0.867(3) & 0.0646(9) & 1.5(1)\\[1.0ex]
\hline\hline  
\end{tabular*}
\label{table:t6} 
\end{table*}

\begin{table*}[h]
\caption{{\bf Supplemental material}) Decomposition of the antipolar structural modification of BiFe$_{0.5}$Sc$_{0.5}$O$_3$ in respect of the symmetrized displacive modes of the parent cubic $Pm\bar{3}m$ perovskite structure (Bi $1b(1/2,1/2,1/2)$, Fe/Sc $1a(0,0,0)$ and O $3d(1/2,0,0)$). The column "Irrep (${\bm k}$)" shows the irreducible representation of the $Pm\bar{3}m$ space group and the arms of the wave vector star involved (for the propagation vectors inside the Brillouin zone, ${\bm -k}$ are not displayed). The column "Order parameter" lists the projections of the reducible order parameter onto the corresponding irreducible subspace (same symbol in different positions indicates equal order parameter components). The columns "Bi, Fe/Sc, O (site irrep)" display amplitudes of the displacive modes and the corresponding point-group symmetry irreps of the local Wyckoff position (in brackets).}
\centering 
\begin{tabular*}{0.98 \textwidth}{@{\extracolsep{\fill}} l l l l l} 
\hline\hline \\ [-1.5ex]
Irrep (${\bm k}$) & Order parameter & Bi (site irrep) &  Fe/Sc(site irrep) & O (site irrep) \\ [1.0ex] 
\hline \\  [-1.5ex]
   & $Pnma$ origin at (${\bm a_p}/2+{\bm b_p}/2)$ &  &  &  \\ [1.0ex]
\hline \\  [-1.5ex]
$X_3^- (0,1/2,0)$ & $(a,0,0)$ & & $-0.11250 (T_{1u})$ & $0.29547 (E_u)$ \\ [1.5ex]

$M_5^- (1/2,0,1/2)$ & $(0,0,a,-a,0,0)$ & $-0.14432 (T_{1u})$ & $-0.06818 (T_{1u})$ & $-0.36365 (E_u)$ \\ [1.5ex]

$R_4^+ (1/2,1/2,1/2)$ & $(\eta ,\eta ,0)$ & & & $-2.61963 (E_u)$\\[1.5ex]

$R_5^+ (1/2,1/2,1/2)$ & $(\delta,-\delta,0)$ & $0.21933 (T1u)$ & &$0.01607 (E_u)$\\[1.5ex]

$\Delta_5 (0,1/4,0)$ & $(0,a,a,0,0,0,0,0,0,0,0,0)$ & $0.15268 (T_{1u})$ & $0.09091 (T_{1u})$ & $-0.10796 (A_{2u})$ \\
& & & & $-0.05114 (E_u)$\\
& & & & $-0.34553 (E_u)$\\ [1.5ex]

$\Lambda_1 (-1/4,1/4,1/4)$ & $(0,0,a,0,0,0,0,a)$ & $-0.02756 (T_{1u})$ & $-0.13176 (T_{1u})$ & $-0.11810 (A_{2u})$ \\
$(1/4,1/4,-1/4)$ & & & & $-0.32476 (E_u)$\\ [1.5ex]

$\Lambda_3 (-1/4,1/4,1/4)$ & $(0,0,0,0,a,-a,-\sqrt{3}a,-\sqrt{3}a,0,0,0,0,\sqrt{3}a,\sqrt{3}a,a,-a)$ & $-0.60405 (T_{1u})$ & $-0.30443 (T_{1u})$ & $-0.02784 (A_{2u})$\\
$(1/4,1/4,-1/4)$ & & & & $-0.37502 (Eu)$\\
& & & & $-0.26900 (E_u)$\\ [1.5ex]

$\Sigma_2 (1/4,0,-1/4)$ & $(0,0,0,0,0,0,a,-a,0,0,0,0)$ & $1.15460 (T_{1u})$ & $0.22728 (T_{1u})$ & $-0.13661 (A_{2u})$ \\
& & & & $1.42231 (E_u)$\\
& & & & $-1.19892 (E_u)$\\ [1.5ex]

$S_4 (1/4,1/2,-1/4)$ & $(0,0,a,-a,0,0,0,0,0,0,0,0)$ & $0.21137 (T_{1u})$ &  & $-0.89999 (E_u)$ \\
& & & & $-0.01705 (E_u)$\\ [1.5ex]

$T_2 (1/2,1/4,1/2)$ & $(0,0,a,a,0,0)$ & & & $0.07387 (A_{2u})$\\[1.5ex]

$T_4 (1/2,1/4,1/2)$ & $(0,0,a,-a,0,0)$ & & & $-0.82390 (E_u)$\\[1.0ex]

\hline \\  [-1.5ex]%
 & $Pnma$ origin at (${\bm a_p}/2+3{\bm b_p}/2$) & & & \\ [0.5ex]
\hline \\  [-1.5ex]

$X_3^- (0,1/2,0)$ & $(a,0,0)$ & & $-0.06429 (T_{1u})$ & $0.25001 (E_u)$ \\ [1.5ex]

$M_5^- (1/2,0,1/2)$ & $(0,0,a,-a,0,0)$ & $-0.14433 (T_{1u})$ & $-0.04546 (T_{1u})$ & $-0.32956 (E_u)$ \\ [1.5ex]

$R_4^+ (1/2,1/2,1/2)$ & $(\eta ,\eta ,0)$ & & & $-2.66455 (E_u)$\\[1.5ex]

$R_5^+ (1/2,1/2,1/2)$ & $(\delta,-\delta,0)$ & $0.20342 (T1u)$ & &$-0.00804 (E_u)$\\[1.5ex]

$\Delta_5 (0,1/4,0)$ & $(a,0,0,-a,0,0,0,0,0,0,0,0)$ & $-0.16071 (T_{1u})$ & $0.03409 (T_{1u})$ & $-0.09091 (A_{2u})$ \\
& & & & $-0.34093 (E_u)$\\
& & & & $-0.24107 (E_u)$\\ [1.5ex]

$\Lambda_1 (-1/4,1/4,1/4)$ & $(0,0,0,a,0,0,-a,0)$ & $-0.00394 (T_{1u})$ & $-0.22269 (T_{1u})$ & $-0.12466 (A_{2u})$ \\
$(1/4,1/4,-1/4)$ & & & & $-0.17166 (E_u)$\\ [1.5ex]

$\Lambda_3 (-1/4,1/4,1/4)$ & $(0,0,0,0,a,a,\frac{1}{\sqrt{3}}a,-\frac{1}{\sqrt{3}}a,0,0,0,0,-\frac{1}{\sqrt{3}}a,\frac{1}{\sqrt{3}}3a,a,a)$ & $-0.59292 (T_{1u})$ & $-0.15747 (T_{1u})$ & $-0.13454 (A_{2u})$\\
$(1/4,1/4,-1/4)$ & & & & $-0.47161 (Eu)$\\
& & & & $0.44944 (E_u)$\\ [1.5ex]

$\Sigma_2 (1/4,0,-1/4)$ & $(0,0,0,0,0,0,a,-a,0,0,0,0)$ & $1.16028 (T_{1u})$ & $0.23865 (T_{1u})$ & $-0.13661 (A_{2u})$ \\
& & & & $1.42232 (E_u)$\\
& & & & $-1.20460 (E_u)$\\ [1.5ex]

$S_4 (1/4,1/2,-1/4)$ & $(0,0,a,-a,0,0,0,0,0,0,0,0)$ & $0.18069 (T_{1u})$ &  & $-0.83571 (E_u)$ \\
& & & & $0.04546 (E_u)$\\ [1.5ex]

$T_2 (1/2,1/4,1/2)$ & $(0,0,a,-a,0,0)$ & & & $-0.32956 (A_{2u})$\\[1.5ex]

$T_4 (1/2,1/4,1/2)$ & $(0,0,a,a,0,0)$ & & & $-0.76140 (E_u)$\\[1.0ex]
\hline\hline 
\end{tabular*}
\label{table:t7} 
\end{table*}

\end{document}